\documentclass[amsmath,amssymb,floatfix,twocolumn,superscriptaddress]{revtex4}
\usepackage{epsf,graphicx}
\usepackage{amssymb}
\usepackage{amsmath}
\usepackage{latexsym,bm,array,amsfonts,multirow}
\usepackage{color}
\usepackage{ulem}
\usepackage[colorlinks=true, letterpaper=true, pdfstartview=FitV, linkcolor=blue, citecolor=blue, urlcolor=blue]{hyperref}

\setcitestyle{super,open={},close={}}

\makeatletter
\renewcommand\@biblabel[1]{#1.}
\makeatother

\begin{document}

\title{Intrinsic constraint on $T_c$ for unconventional superconductivity}
\author{Qiong Qin}
\affiliation{Beijing National Laboratory for Condensed Matter Physics and Institute of
	Physics, Chinese Academy of Sciences, Beijing 100190, China}
\affiliation{University of Chinese Academy of Sciences, Beijing 100049, China}
\author{Yi-feng Yang}
\email[]{yifeng@iphy.ac.cn}
\affiliation{Beijing National Laboratory for Condensed Matter Physics and Institute of
Physics, Chinese Academy of Sciences, Beijing 100190, China}
\affiliation{University of Chinese Academy of Sciences, Beijing 100049, China}
\affiliation{Songshan Lake Materials Laboratory, Dongguan, Guangdong 523808, China}
%\date{\today}

\begin{abstract}
Can room temperature superconductivity be achieved in correlated materials under ambient pressure? Our answer to this billion-dollar question is probably no, at least for realistic models within the current theoretical framework. This is shown by our systematic simulations on the pairing instability of some effective models for two-dimensional superconductivity. For a square lattice model with nearest-neighbour pairing, we find a plaquette state formed of weakly-connected $2\times2$ blocks for sufficiently large pairing interaction. The superconductivity is suppressed on both sides away from its melting quantum critical point. Thus, the magnitude of $T_c$ is constrained by the plaquette state for the $d$-wave superconductivity, in resemblance of other competing orders. We then extend our simulations to a variety of effective models covering nearest-neighbour or onsite pairings, single layer or two-layer structures, intralayer or interlayer pairings, and find an intrinsic maximum of the ratio $T_c/J\approx 0.04-0.07$, where $J$ is the pairing interaction, given by the onsite attractive interaction in the attractive Hubbard model or the exchange interaction in the repulsive Hubbard model. Our results agree well with previous quantum Monte Carlo simulations for the attractive Hubbard model. Comparison with existing experiments supports this constraint in cuprate, iron-based, nickelate, and heavy fermion superconductors, despite that these compounds are so complicated well beyond our simplified models. As a result, the known families of unconventional superconductivity, possibly except the infinite-layer nickelates, seem to almost exhaust their potentials in reaching the maximal $T_c$ allowed by their respective $J$, while achieving room temperature superconductor would require a much larger $J$ beyond 400-700 meV, which seems unrealistic in existing correlated materials and hence demands novel pairing mechanisms. The agreement also implies some deep underlying principles of the constraint that urge for a more rigorous theoretical understanding.
\end{abstract}

\maketitle

\noindent
\textbf{Introduction}\\
Despite the century-long pursuit of high-temperature superconductors, the possible existence of a theoretical upper limit to their transition temperature ($T_c$) under ambient pressure remains unsettled \cite{Zhang2023,Paiva2010,Hazra2019,Hofmann2022,Esterlis2018}. Both mean-field and weak-coupling Eliashberg theories \cite{Monthoux1992b} predict an artificial $T_c$ that grows continuously with increasing pairing interaction, while experiments often find superconducting domes with maximum $T_c$ near the phase boundaries of some long- or short-range orders associated with spin, charge, orbital, or structural degrees of freedom \cite{Mathur1998,Monthoux2007,Norman2011,Kivelson2006,Ganin2008,Wu2014,Seo2015,Chen2021,Gruner2017,Yu2021,Klein2019,Profe2024}. The dome implies a dual role of the  many-body interaction \cite{Balseiro1979}, which may not only provide the pairing glue but also induce competing orders that constrain the maximum $T_c$. However, they are mostly external factors associated with instabilities of other channels. One may wonder if any intrinsic constraint on $T_c$ may exist owing solely to the pairing instability.

Important lessons may be learned from cuprate high-temperature superconductors in the underdoped region, where strong pairing interactions relative to the renormalized effective quasiparticle hopping parameters favor short-range electron pairs \cite{Sobirey2022a} that in some literatures are thought to form already at high temperatures but only become superconducting when a (quasi-)long-range phase coherence is developed \cite{Keimer2015,Emery1995}. This raises a few general questions: What is the true strong-coupling limit of the pairing state? How is this strong-coupling state related to the high-temperature superconductivity? Would it put any intrinsic constraint on the maximal value of $T_c$? Since the absolute magnitude of $T_c$ is determined by certain basic energy scale, such as the pairing interaction $J$, the question of maximum $T_c$ turns into the question of their maximum dimensionless ratio $T_c/J$. To address these important issues and gain insights into possible intrinsic constraints on $T_c$, we propose here to discard instabilities from all other channels such as magnetic or charge orders and focus only on the pairing instability, since all other instabilities are expected to compete with the superconductivity and further suppress $T_c$. Their contributions to the pairing can all be included phenomenologically in a pairing interaction term. 

Theoretically, one may derive various ratios with respect to other measurable energy scales such as the Fermi energy and the superfluid density. However, it has been shown that these constraints may be violated in artificial models \cite{Hofmann2022}, or even in real materials \cite{Tian2023Nature}, thus preventing a useful bound for constraining $T_c$. To avoid such complication, instead of deriving a model-independent constraint, we first restrict ourselves to a minimal effective model that is most relevant in real correlated superconductors and includes only the quasiparticle hopping and nearest-neighbour spin-singlet pairing interaction. For the one band model on a square lattice, we find a plaquette state in the strong-coupling limit that breaks both the translational and time-reversal symmetries and exhibits unusual spectral properties with a pseudogap or insulating-like normal state. The $d$-wave superconductivity emerges as the plaquettes melt and short-range electron pairs get mobilized to attain long-distance phase coherence at a reduced pairing interaction. A tentative phase diagram is then constructed where $T_c/t$ reaches its maximum at the plaquette quantum critical point (QCP), resembling those often observed in experiments with other competing orders. This suggests some intrinsic constraints that prevent $T_c$ from exhausting all kinetic or pairing energies in order to achieve a delicate balance between pairing and phase coherence. We then extend the calculations to more general models with either nearest-neighbour or onsite pairings, single layer or two-layer structures, intralayer or interlayer pairings, and obtain a maximum $T_c/J\approx 0.04-0.07$. A close examination of existing experiments in known unconventional superconductors, including cuprate, iron-based, nickelate, and heavy fermion superconductors, seems to quite universally support the obtained ratio, indicating that these families, possibly except the infinite-layer nickelates, have almost reached their maximum $T_c$ allowed by their respective spin exchange interactions. A room-temperature superconductor would then require a much larger pairing interaction beyond 400-700 meV within the current theoretical framework, which seems unrealistic from a single mechanism in correlated electron systems under ambient pressure. Our work therefore provides a useful criterion that may help to avoid futile efforts in exploring high-temperature superconductors along wrong directions. It also points out the necessity of new pairing mechanisms, possibly combining different pairing interactions, in order to achieve the room-temperature superconductivity.

\begin{figure*}[t]
\centering
\includegraphics[width=0.6\linewidth]{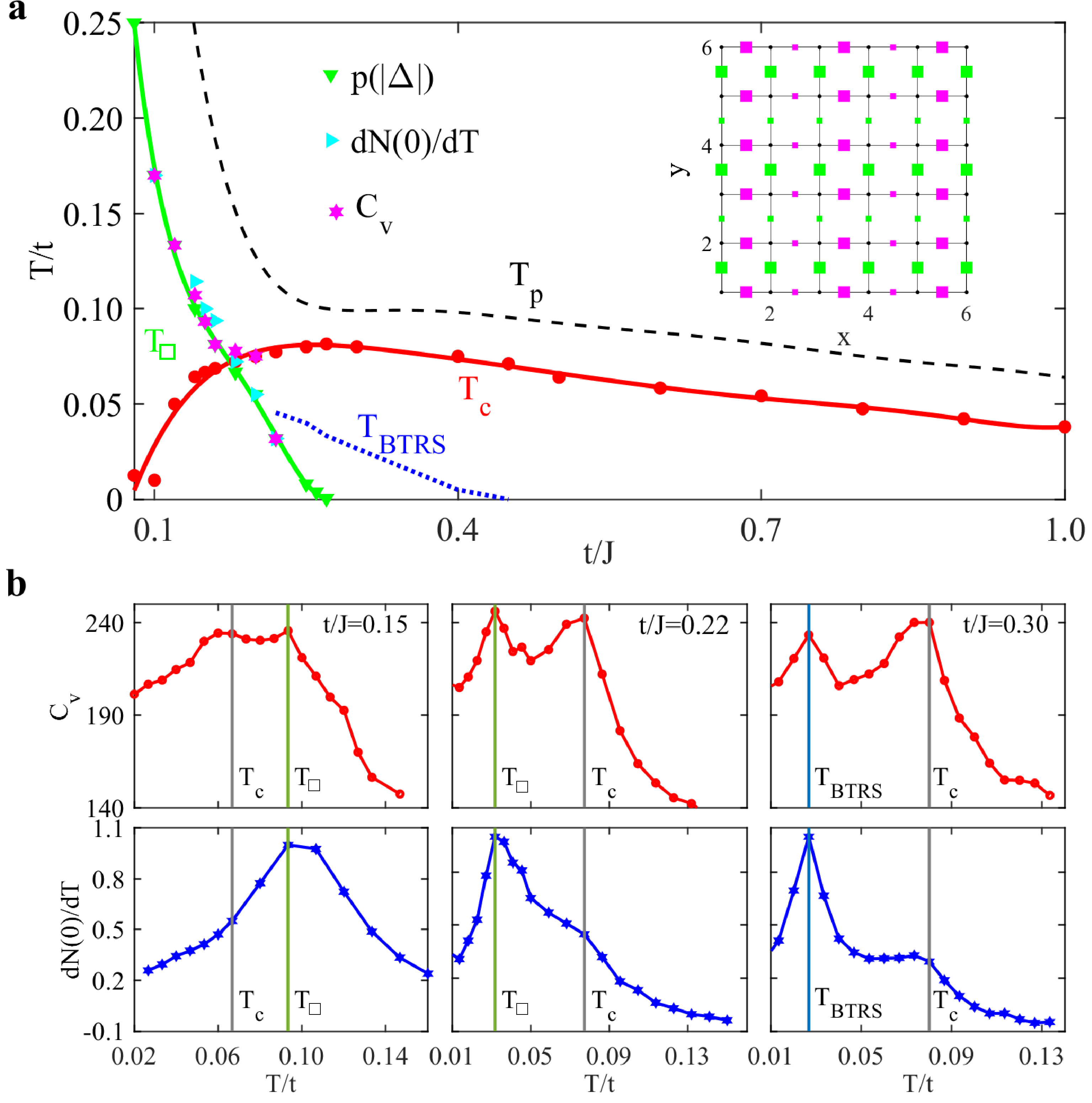}
\caption{\textbf{Theoretical phase diagram of our minimal effective model on the square lattice.} \textbf{a} The $T/t-t/J$ phase diagram, showing the onset temperature $T_p$ of pseudogap-like behavior determined from the suppression of the quasiparticle density of states at the Fermi energy $N(0)$, the plaquette transition temperature $T_{\scriptscriptstyle \square}$ from the pairing field amplitude distribution $p(|\Delta|)$ and the peaks in the specific heat $C_{\rm v}$ and the temperature derivative of the quasiparticle density of states $dN(0)/dT$, the superconducting transition temperature $T_c$ from the long-distance phase coherence based on the phase mutual information and the properties of the Berezinskii-Kosterlitz-Thouless (BKT) transition for two-dimensional superconductivity, and the temperature $T_{\rm BTRS}$ for superconductivity with broken time-reversal symmetry from the deviation of the pairing field phases along $x$ and $y$ bonds attached to the same site. The inset shows a typical configuration of the pairing field inside the plaquette state at low temperatures, where the size of the symbols represents the amplitude $|\Delta_{ij}|$ and the color denotes the sign of the phase $\theta_{ij}$. Note that for small $t/J$, $T_{\rm BTRS}$ extends to the plaquette transition $T_{\scriptscriptstyle \square}$, which breaks simultaneously the time-reversal symmetry and the translational symmetry and coexists with a bond charge order (see Methods). \textbf{b} Temperature evolution of $C_{\rm v}$ and  $dN(0)/dT$ for $t/J=0.15$, 0.22, 0.30, showing peaks or shoulders at $T_{\scriptscriptstyle \square}$, $T_c$, and $T_{\rm BTRS}$.}
\label{fig1}
\end{figure*}

\vspace{20pt}
\noindent
\textbf{Results}\\
\noindent\textbf{Model and method}\\
We start by first considering an effective one-band $t$-$J$ type model on the square lattice, which will later be extended to more general cases (see Methods). The model Hamiltonian is written as:
	\begin{eqnarray}
		H=-\sum_{ij,\sigma}t_{ij}d_{i\sigma}^{\dagger}d_{j\sigma}-\mu\sum_{i\sigma}d_{i\sigma}^{\dagger}d_{i\sigma}-J\sum_{\langle ij\rangle}\psi_{ij}^{\dagger}\psi_{ij},
	\end{eqnarray}
	where $t_{ij}$ is the renormalized quasiparticle hopping parameter, $\mu$ is the chemical potential, and the pairing interaction is written in terms of the spin-singlet operator $\psi_{ij}=\frac{1}{\sqrt{2}}(d_{i\downarrow}d_{j\uparrow}-d_{i\uparrow}d_{j\downarrow})$ between nearest-neighbour sites, where $d_{i\sigma}(d_{i\sigma}^{\dagger})$ is the annihilation (creation) operator of the quasiparticles to be paired. For the exchange mechanism, the pairing interaction may be induced by the nearest-neighbour antiferromagnetic interaction and the attractive charge density interaction. As shown in Methods, $J$ is then simply the nearest-neighbor exchange interaction, whose importance in cuprates has been justified in numerous experiments \cite{Ofer2006a,Tacon2011,Dean2013,Ruan2016,Wang2023a}. To study the superconductivity, we introduce the complex auxiliary fields $\Delta_{ij}$ to decouple the pairing term \cite{Coleman2015}:
	\begin{equation}
		-\psi_{ij}^{\dagger}\psi_{ij} \rightarrow \frac{\sqrt{2}}{J}\left(\bar{\Delta}_{ij}\psi_{ij}+\psi_{ij}^{\dagger}\Delta_{ij} \right)+\frac{2|\Delta_{ij}|^2}{J^2}.
	\end{equation}
	To avoid the negative sign problem, we assume a static approximation, $\Delta_{ij}(\tau) \rightarrow \Delta_{ij}=|\Delta_{ij}|e^{i\theta_{ij}}$, and employ the auxiliary field Monte Carlo approach \cite{Mayr2005a,Dubi2007,Pasrija2016,Dong2021a,Mukherjee2014,Atkinson2012,Zhong2011,Qin2023PRB,Dong2022PRB}. Following the standard procedure, we integrate out the fermionic degrees of freedom and simulate the final effective action only of the pairing fields by the Metropolis algorithm \cite{Qin2023PRB}. 
	
	This method ignores dynamic fluctuations of the pairing fields but takes full consideration of their spatial and thermal fluctuations. We can investigate phase correlations of the pairing and determine $T_c$ based on long-distance phase coherence rather than the BCS-type mean-field transition. The validity of our method in estimating the maximum $T_c$ has been verified in the recently-discovered bilayer and trilayer nickelate superconductors \cite{Qin2023a,Qin2024} and by its consistency with the rigorous Quantum Monte Carlo simulations for the attractive Hubbard model \cite{Paiva2010}. However, its applications in analyzing certain dynamical properties of the pairing fields are limited. To maximize $T_c$, we have also ignored all other instabilities outside the pairing channel. The effect of the Gutzwiller constraint is approximated by treating $t_{ij}$ as free tuning parameters.
	
For numerical calculations, we consider a $10\times 10$ square lattice with periodic boundary conditions and include only the nearest-neighbour hopping $t$ and the next-nearest-neighbour hopping $t^\prime=-0.45 t$ as in cuprate high-temperature superconductors \cite{Carbotte1994,Monthoux2001}. The chemical potential is fixed to $\mu=-1.4 t$. The presented results have been examined and found qualitatively consistent for other values of the parameters or on a larger lattice. A twisted boundary condition is used for spectral calculations \cite{Li2018PRL}.\\

\begin{figure*}[t]
\centering
\includegraphics[width=0.6 \linewidth]{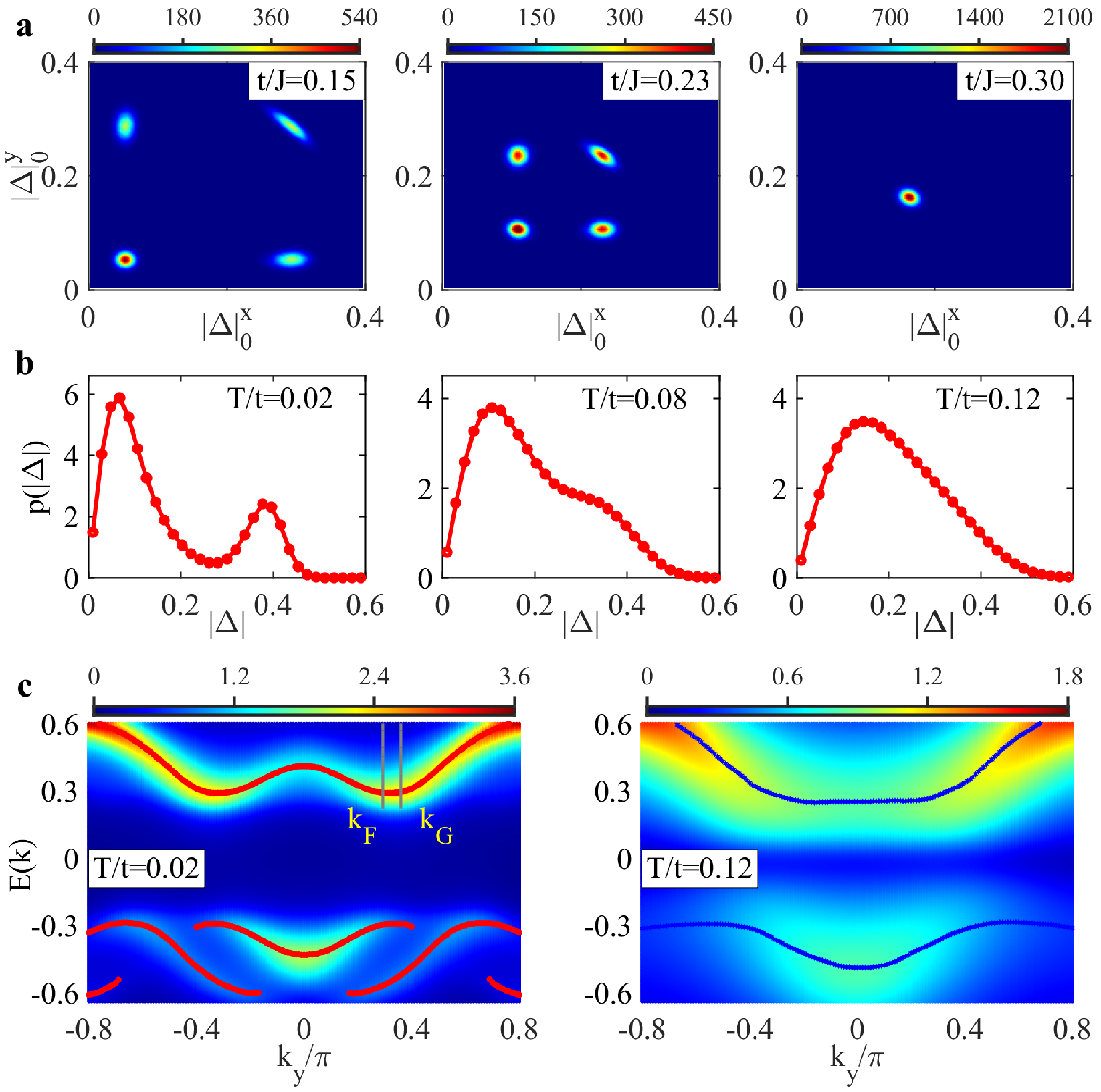}
\caption{ \textbf{Properties of the plaquette state at strong coupling.} \textbf{a} The joint distribution function $p(|\Delta|_{\bm{0}}^x,~|\Delta|_{\bm{0}}^y)$ of the pairing field amplitudes $|\Delta_{\bm{0}}^{x}|$ and $|\Delta_{\bm{0}}^{y}|$ along $x$ and $y$ directions attached to the same site $\bm{0}$ for $t/J=0.15$, 0.23, 0.30 at a very low temperature $T/J=0.0001$. \textbf{b} Evolution of the marginal distribution $p(|\Delta|)$ of the pairing amplitude on all bonds with temperature for $t/J=0.15$. \textbf{c} Comparison of the energy-momentum dependent spectral function and extracted dispersions (solid lines) at $k_x/\pi=0.44$ at low and high temperatures for $t/J=0.15$. The grey vertical lines mark the Fermi vector $k_{\rm F}$ that clearly differs from the wave vector $k_{\rm G}$ where the dispersions bend backwards.}
	\label{fig2}
\end{figure*}

\noindent\textbf{Theoretical phase diagram}\\
Figure~\ref{fig1}\textbf{a} shows a typical theoretical phase diagram for the one-band square lattice model, where we have intentionally plot $T_c/t$ against $t/J$. A nonuniform plaquette state emerges at sufficiently strong paring interaction formed of $2\times2$ blocks induced by high-order pair hopping in the effective action of the pairing fields after integrating out the electron degrees of freedom. A typical pairing configuration of the plaquette state is given in the inset of Fig.~\ref{fig1}. The paring amplitudes are relatively stronger on internal bonds of the $2\times2$ plaquettes and weaker on their links. As will be discussed later and in the Methods, the plaquette state simultaneously breaks the lattice translational symmetry and the time-reversal symmetry and coexists with a bond charge order, though the electron density remains uniform on all sites. Its transition temperature $T_{\scriptscriptstyle \square}$ decreases with increasing $t/J$ and diminishes at the QCP ($t/J\approx 0.27$), where the plaquettes melt completely and uniform superconductivity emerges with a maximum $T_c/t\approx0.08$ for the chosen parameters (see Methods). Tuning the next-nearest-neighbour hopping and the chemical potential may slightly change the ratio and the location of the QCP, but does not alter the qualitative physics. Inside the plaquette state, the superconductivity is also spatially modulated and its $T_c$ is greatly reduced as the pairing interaction further increases. The nonmonotonic evolution of $T_c$ resembles typical phase diagrams observed in many unconventional superconductors with other competing orders such as long-range magnetism, charge density wave, or nematicity \cite{Mathur1998,Monthoux2007,Norman2011,Kivelson2006,Ganin2008,Wu2014,Seo2015,Chen2021,Gruner2017,Yu2021}. However, the plaquette state reflects the internal instability in the pairing channel that constrains the magnitude of $T_c$ for the $d$-wave superconductivity. Near the plaquette QCP, the superconductivity also breaks the time-reversal symmetry below $T_{\rm BTRS}$. As $t/J$ decreases, the $T_{\rm BTRS}$ transition line merges with the plaquette transition $T_{\scriptscriptstyle \square}$, marking the simultaneous time-reversal symmetry breaking of the plaquette state. At high temperatures, the normal state exhibits pseudogap-like behavior whose onset temperature $T_p$ follows closely the variation of $T_c$ or $T_{\scriptscriptstyle \square}$ \cite{Armitage2010,Jang2023} determined from the specific heat $C_{\rm v}$ or the temperature derivative of the quasiparticle density of states at the Fermi energy $dN(0)/dT$. As shown in Fig. \ref{fig1}\textbf{b}, we find peaks in the specific heat for all transitions at $T_{\scriptscriptstyle \square}$, $T_{c}$, and $T_{\rm BTRS}$, while in $dN(0)/dT$ the feature at $T_c$ is greatly suppressed for $t/J<0.27$. Here and after, $J$ is set as the energy unit if not explicitly noted.\\

\begin{figure*}[t]
\centering
\includegraphics[width=0.6 \linewidth]{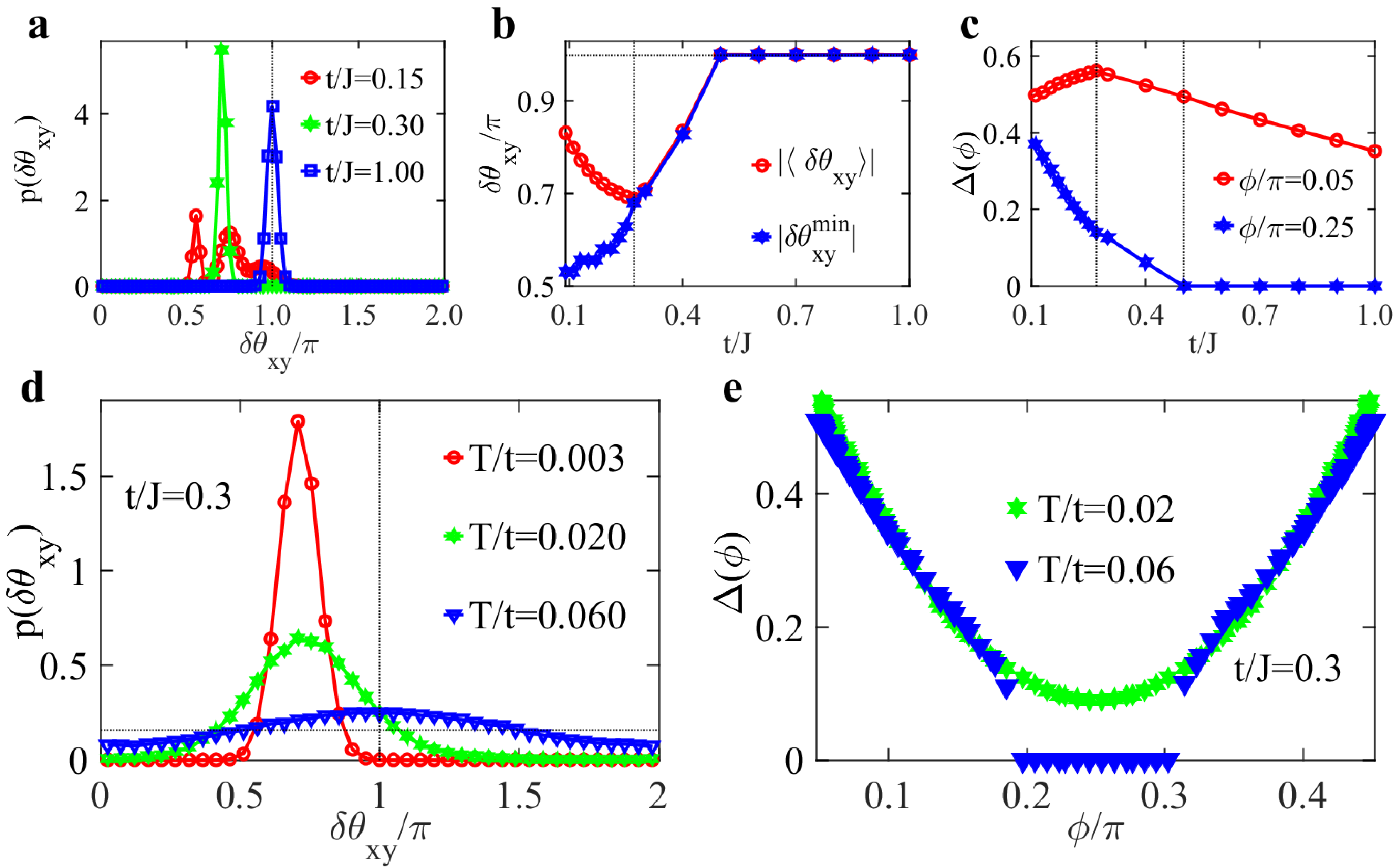}
\caption{\textbf{Two-stage quantum phase transition and time-reversal symmetry breaking.} \textbf{a} The probabilistic distribution of the phase difference along $x$ and $y$ directions $p(\delta \theta_{xy})=p(\theta_{\bm{0}}^x-\theta_{\bm{0}}^y)$ for different values of $t/J$. \textbf{b} Evolution of the average phase difference $|\langle \delta \theta_{xy}\rangle|$ and the minimum phase difference $|\delta \theta_{xy}^{\rm min}|$ determined by the peak positions as functions of $t/J$. The vertical line marks the plaquette QCP at $t/J=0.27$. \textbf{c} Comparison of the gap function $\Delta(\phi)$ near nodal and antinodal directions as functions of $t/J$, determined by the position of the positive-energy peak in the spectral function $A(\phi,\omega)$. The temperature is $T/J=0.0001$. \textbf{d} Temperature evolution of $p(\delta \theta_{xy})$ at $t/J=0.3$ in the intermediate phase. \textbf{e} The angle-dependent gap function $\Delta(\phi)$ for different temperatures at $t/J=0.3$, showing the evolution from a full gap at low temperatures to a partial gap at high temperatures. }
	\label{fig3}
\end{figure*}

\noindent\textbf{Plaquette state at strong coupling}\\
The plaquette state and its phase transition may be seen in the joint distribution $p(|\Delta|_{\bm{0}}^x,~|\Delta|_{\bm{0}}^y)$ of the paring amplitudes along the $x$ and $y$ directions attached to the same site $\bm{0}$ or the marginal distribution $p(|\Delta|)$ of the pairing field amplitudes on all bonds. As shown in Fig.~\ref{fig2}\textbf{a}, $p(|\Delta|_{\bm{0}}^x,~|\Delta|_{\bm{0}}^y)$ at low temperatures displays a four-point structure due to the nonuniform pairing configurations. As $t/J$ increases, the four points gradually shrink into a single point, where the translational symmetry is recovered and the plaquette state melts into the uniform superconductivity. Correspondingly, the amplitude distribution $p(|\Delta|)$ also contains two peaks in the plaquette state. As shown in Fig. \ref{fig2}\textbf{b} for $t/J=0.15$, these peaks get gradually broadened with increasing temperature and merge into a single peak above $T_{\scriptscriptstyle \square}$.

At sufficiently low temperatures, the plaquette state may develop long-distance phase coherence to form the superconductivity but exhibits unusual spectral features due to the nonuniform spatial distribution of the pairing amplitudes. As shown in Fig. \ref{fig2}\textbf{c} for $t/J=0.15$, its momentum-energy dependent spectral function at negative energies splits into two sets of dispersions. One dispersion resembles that of uniform superconductivity, but its back-bending vector $k_{\rm G}$ differs consistently from the Fermi vector $k_{\rm F}$, which has also been observed experimentally for possible pair density wave (PDW) state \cite{Agterberg2008,Lee2014,He2011a}. At high temperatures, the two dispersions recombine into a single curve pointing upwards even in the normal state. The gap indicates a pseudogap or insulating-like phase due to the large nearest-neighbour pairing interaction. This suggests that the normal state may also undergo a metal-insulator transition as $t/J$ decrease, a phenomenon observed in cuprate superconductors under high pressure but unexplained \cite{Zhou2022a}. At intermediate temperature $T_c<T<T_{\scriptscriptstyle \square}$, the superconducting phase coherence is lost and the plaquette state with broken time-reversal symmetry is in a sense similar to the fermionic quadrupling phase proposed earlier in experiment \cite{Grinenko2021a,Shipulin2023}.\\

 \begin{figure*}[t]
\centering
 \includegraphics[width=0.6 \linewidth]{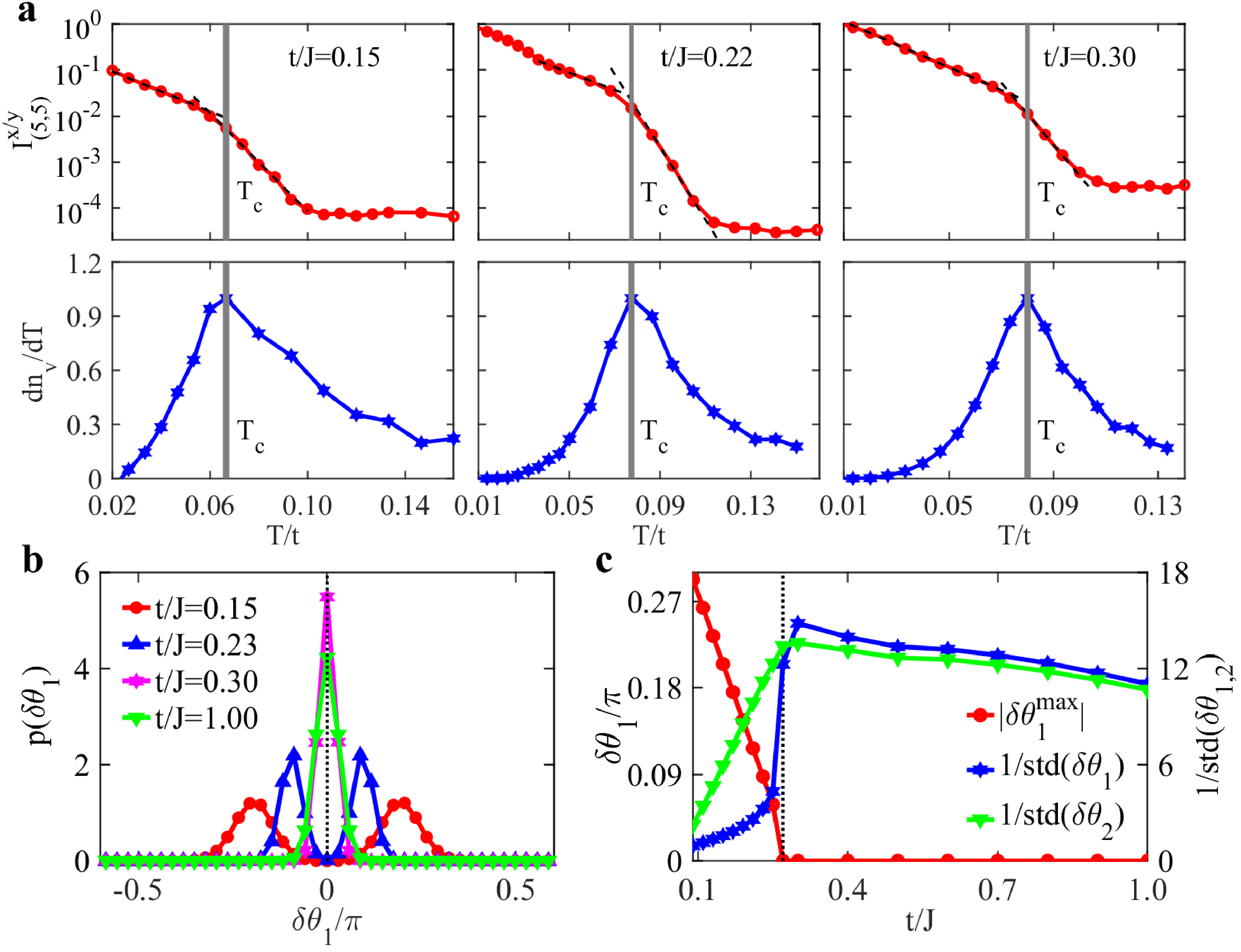}
 \caption{\textbf{Superconducting phase coherence.} \textbf{a} The mutual information between two pairing field phases $\theta_{(0,0)}^{x/y}$ and $\theta_{(5,5)}^{x/y}$ of the distance $(5,5)$ and the normalized numerical derivatives of the vortex number $dn_{\rm v}/dT$ with temperature for $t/J=0.15,0.22,0.30$, respectively. The vertical lines show the extracted $T_c$. \textbf{b} The distribution $p(\delta \theta_1)$ for different hopping at $T/J=0.0001$, where $\delta \theta_1$ is the phase difference between nearest-neighbour pairing fields  $\delta \theta_1=\theta_{\bm{0}}^{x/y}-\theta_{(1,0)/(0,1))}^{x/y}$. \textbf{c}  The peak position in $p(\delta \theta_1)$ and the inverse of the fluctuation ${\rm std}(\delta\theta_i)=\sqrt{\langle(\delta\theta_i)^2\rangle}$, where $\delta\theta_i$ is the phase difference between two nearest-neighbour ($i=1$) or next-nearest-neighbour ($i=2$) bonds along $x$ or $y$ directions. The two behave similarly for uniform superconductivity but differ in the plaquette state.}
  	\label{fig4}
  \end{figure*}

\noindent\textbf{Time-reversal symmetry breaking}\\
The time-reversal symmetry breaking may be seen from the probabilistic distribution $p(\delta \theta_{xy})$ of the phase difference $\delta \theta_{xy}=\theta_{\bm{0}}^x-\theta_{\bm{0}}^y$ of the pairing fields along the $x$ and $y$ directions. The results are shown in Fig. \ref{fig3}\textbf{a} for three different values of $t/J$ at very low temperature. For small $t/J=0.15$ in the plaquette state, the existence of multiple peaks mark the phase difference on different bonds. These peaks develop into a two-peak structure at higher temperatures and then merge into a single peak at $\delta\theta_{xy}=\pi$ above $T_{\scriptscriptstyle \square}$ (see Methods). For large $t/J=1.0$, there exists a single maximum around $\delta\theta_{xy}/\pi=1$, which signals the uniform $d$-wave superconductivity with opposite sign of the pairing field along the $x$ and $y$ directions. Quite unexpectedly, for $t/J=0.3$, we still have a single peak but its position deviates from $\delta\theta_{xy}/\pi=1$. To see such a variation more clearly, Fig. \ref{fig3}\textbf{b} plots the average deviation ($|\langle \delta \theta_{xy}\rangle|$) and the smallest deviation ($|\delta \theta_{xy}^{\rm min}|$) of the peak positions. While $|\langle \delta \theta_{xy}\rangle|$ evovles nonmonotonically and reaches a minimum at the plaquette QCP, $|\delta \theta_{xy}^{\rm min}|$ keeps increasing with $t/J$. Interestingly, the two quantities become equal beyond the plaquette QCP but only approach $\pi$ at a much larger $t/J\approx 0.5$. 

Under time-reversal operation, the phase of the pairing field changes sign so that $\delta\theta_{xy} \rightarrow -\delta\theta_{xy}$ (mod 2$\pi$). Thus, the deviation of the peak position from $\pi$ around $t/J=0.15$ indicates that the time-reversal symmetry is broken inside the whole plaquette state. For $t/J=0.3$, it marks an intermediate region of uniform superconductivity that breaks the time-reversal symmetry, with the gap function $\Delta_{\bm k}\propto \cos({\bm{k}}_x)+e^{-i\delta\theta_{xy}}\cos({\bm{k}}_y)\propto \Delta^d_{\bm k} -i\cot\frac{\delta\theta_{xy}}{2}\Delta^s_{\bm k}$, representing $d+is$ pairing with a nodeless gap. Here $\Delta^d_{\bm k}=\cos({\bm{k}}_x)-\cos({\bm{k}}_y)$ is the $d$-wave component and $\Delta^s_{\bm k}=\cos({\bm{k}}_x)+\cos({\bm{k}}_y)$ denotes an extended $s$-wave component from the nearest-neighbour pairing interaction. The onsite pairing is not included due to the strong Coulomb repulsion. We have therefore a two-stage transition from the plaquette to the uniform $d$-wave superconductivity, with an intermediate region that recovers the translational symmetry but still breaks the time-reversal symmetry. Similar $d+is$ pairing may have been found under certain conditions in twisted double-layer cuprates \cite{Can2021b} and infinite-layer nickelates \cite{Wang2020PRB,Ji2023}. In the latter case, it arises from the interplay of Kondo and superexchange interactions \cite{Zhang2020a}. Here it is associated with the quasiparticle hopping, $i \rightarrow i+\hat{x} \rightarrow i+\hat{x}+\hat{y} \rightarrow i+\hat{y} \rightarrow i$. Integrating out the electron degrees of freedom leads to a term like ${\rm Re}(\Delta_{i,i+\hat{x}}\Delta_{i+\hat{y},i+\hat{x}+\hat{y}}\Delta^*_{i,i+\hat{y}}\Delta_{i+\hat{x},i+\hat{x}+\hat{y}}^*) \rightarrow {\rm Re}(\Delta_x^2\Delta_y^*{}^2)\propto \cos(2\delta\theta_{xy})$, while the second order hopping process such as $i+\hat{x} \rightarrow i\rightarrow i+\hat{y}$ contributes a term ${\rm Re}(\Delta_x\Delta_y^*) \propto \cos\delta\theta_{xy}$. Their combined free energy may be minimized at $\delta\theta_{xy}$ away from 0 and $\pi$ \cite{Tesanovic2008}. Thus, time-reversal symmetry breaking represents an intrinsic tendency of the superconductivity with nearest-neighbour pairing at strong coupling, where the normal state is no longer a Fermi liquid.

To further confirm the two-stage transition, Fig. \ref{fig3}\textbf{c} plots the gap function $\Delta(\phi)$ with $t/J$ near the nodal and antinodal directions in the momentum space deduced from the spectral function. The gap near the antinode is always finite, but varies nonmonotonically with a maximum at the plaquette QCP $t/J=0.27$, in good correspondence with the maximum $T_c$. By contrast, the gap near the nodal direction decreases continuously and only diminishes at $t/J\approx0.5$, confirming a full gap for $0.27\le t/J\le 0.5$ consistent with the above phase analysis. The transition temperature $T_{\rm BTRS}$ of the $d+is$ phase may also be extracted from the temperature evolution of $p(\delta \theta_{xy})$. As shown in Fig.~\ref{fig3}\textbf{d} for $t/J=0.3$, the peak in $p(\delta \theta_{xy})$ gets broadened and moves gradually to $\delta\theta_{xy}=\pi$ as the temperature increases across $T_{\rm BTRS}$. The angle-dependent gap functions are given in Fig.~\ref{fig3}\textbf{e}, showing a fully gapped $d+is$ pairing state and a nodal $d$-wave pairing state below and above $T_{\rm BTRS}$, respectively. Note that the higher-temperature $d$-wave gap contains a finite gapless region on the Fermi surface, which has also been observed previously in some experiments \cite{Vishik2012}.\\

\noindent\textbf{Superconducting phase coherence}\\
The superconducting transition is determined from the phase mutual information $I^{x/y}_{\bm R}$ of the pairing fields as well as the vortex number $n_{\rm v}$ (see Methods) \cite{Qin2023PRB,Qin2023a}. Figure \ref{fig4}\textbf{a} shows the semilog plot of the phase mutual information between two bonds of the largest distance ${\bm R}=(5,5)$ for $t/J=0.15$, 0.22, 0.30 on the 10$\times$10 lattice. We find a slope change at low temperature, marking the establishment of long-distance phase coherence of the pairing fields. The slope change at higher temperature is associated with the onset of the spatial phase correlation, which has a temperature scale in rough agreement with $T_p$ for $t/J>0.27$ in Fig. \ref{fig1} and is therefore responsible for the pseudogap above the superconducting $T_c$.

The low-temperature transition coincides with the peak position of $dn_{\rm v}/dT$ also plotted in Fig. \ref{fig4}\textbf{a}. The maximum of $dn_{\rm v}/dT$ implies a rapid development of the vortex number $n_{\rm v}$ with increasing temperature, which is a characteristic feature of the Berezinskii-Kosterlitz-Thouless (BKT)  transition for two-dimensional superconductivity \cite{Berezinskii1972, Kosterlitz1973, Kosterlitz1974}. We thus identify this transition as the superconducting transition. The value of $T_c$ is examined for other lattice size and found to vary only slightly, confirming the robustness of our qualitative conclusions.

The final phase diagram is already discussed in Fig. \ref{fig1}\textbf{a}, showing nonmonotonic variation of $T_c/t$ with $t/J$ and a maximum at the plaquette QCP. The overall evolution of $T_c$ may be understood from the phase difference of the pairing fields on neighbouring bonds. Figure \ref{fig4}\textbf{b} plots the probabilistic distribution $p(\delta \theta_1)$ of $\delta \theta_1=\theta_{\bm{0}}^{x/y}-\theta_{(1,0)/(0,1)}^{x/y}$. We find two symmetric peaks around zero in the plaquette state and a single peak in the uniform superconducting state. Interestingly, as shown in Fig.~\ref{fig4}\textbf{c}, while the peak position $|\delta \theta_1^{\rm max}|$ decreases gradually and diminishes above $t/J=0.27$, the inverse of its fluctuation, as well as that between next-nearest-neighbour bonds, also varies nonmontonically with $t/J$ and exhibits a maximum near the plaquette QCP, in good correspondence with the evolution of $T_c/t$. This coincidence is unexpected at first glance but easy to understand, since a smaller fluctuation of $\delta\theta_1$ around zero indicates a larger phase stiffness of the pairing fields on neighbouring bonds, thus favoring larger superfluid density and $T_c$. Theoretically, this is usually described by the free energy \cite{Coleman2015}, $F=\frac{\rho_{\rm s}}{2}\int_x (\delta \theta)^2$, such that the phase fluctuation $\langle(\delta\theta)^2\rangle$ is inversely related to the superfluid density $\rho_{\rm s}$. This explains our observed correlation between the fluctuation of the phase difference and the magnitude of $T_c$ in Fig. \ref{fig4}\textbf{c}.\\

\noindent\textbf{Discussion on the plaquette state}\\
The plaquette state may also have other exotic properties detectable in experiments. For example, pairing field modulation may affect local spin susceptibility \cite{Choubey2020,Kinjo2022a} and cause some spin resonance mode \cite{Stock2008a}. In fact, the plaquette state shares many similarities with the supersolid phase realized in dipolar cold atoms \cite{Saccani2012,Natale2019,Guo2019,Ilzhofer2021,Tanzi2019}. Both break translational symmetry and $U(1)$ phase symmetry at zero temperature. Similar to the plaquette state, the microscopic configurations of supersolid consist of weakly connected droplets. Both occupy an intermediate region of their respective phase diagram: the plaquette state occurs between the uniform superconductivity and a disordered phase of coexisting plaquettes and dimers for extremely large pairing interaction, while the supersolid exists between the superfluid phase and an incoherent droplet solid. Given these similarities, one may anticipate that vortices may exist in the supersolid phase, while two modes with different dispersions for some dynamic structure factor observed in supersolid \cite{Natale2019} may also emerge in the plaquette state.

Though the Bose-Einstein condensation (BEC) \cite{Chen2022b} has traditionally been argued to be the strong coupling limit of the superconductivity, our results question its naive extension to unconventional superconductor that cannot be described by the local $s$-wave pairing with onsite attractive interaction. In particular, for unconventional superconductors with strong onsite Coulomb repulsion such as the cuprates, onsite $s$-wave pairing is generally unfavored and the pairs tend to occupy different sites. As a result, short-range pairing emerges for nearest-neighbour spin exchange interaction and, at strong coupling, may cause plaquette states that break the lattice translational symmetry. This differs from the local two-particle bound state typical of the BEC. On the other hand, our derived plaquette state does share some similarities with the BEC, such as the U-shaped density of states near the Fermi energy, the flat dispersion around $k_x=0$, and the pseudogap in the normal state at high temperatures.

Our proposed plaquette state is different from the widely-studied PDW state \cite{Agterberg2008,Agterberg2020,Lee2014,Berg2009,Wu2024}, even though both exhibit real-space modulation of the pairing fields. While the PDW may generally lead to a charge density wave, the plaquette state breaks simultaneously the time-reversal symmetry and the translational symmetry and coexists with a bond charge order \cite{Balseiro1979}. The PDW is by far only found experimentally in superconducting region \cite{Chen2021g,Zhao2023,Du2020} and might arise theoretically from the interplay of magnetism and superconductivity \cite{Chen2023a}, while the plaquette state proposed here represents an intrinsic pairing instability at strong coupling.\\

\noindent\textbf{Relevance to the cuprates}  \\
We note again that $t$ is the effective hopping parameter of the quasiparticles, which should already take account of the Gutzwiller constraint. It is a small number proportional to the hole doping in underdoped cuprates, and reaches about $100-200$ meV in overdoped cuprates as indicated by ARPES measurements \cite{Chen2022}. By contrast, the exchange interaction $J$ is almost doping-independent and roughly $100-200$ meV as revealed by RIXS experiments \cite{Ofer2006a,Tacon2011,Dean2013}. Hence, our choice of the $t/J$ range is reasonable according to these experiments. Numerically, our results cover the strong to weak coupling regions of the superconducting pairing, and provide useful information on the maximum $T_c$ and the plaquette instability. Many of our findings are in good correspondence with experimental observations \cite{Ye2023a,Li2023,Zhou2022a,Kaminski2002,He2011a}. In particular, the plaquette structure might be closely related to the $4\times 4$ structure observed in recent STM measurements on underdoped cuprates  \cite{Ye2023a,Li2023}, suggesting that Cooper pairs may first form within these local structures and only achieve long-range phase coherence as the doping reaches a certain threshold \cite{Ye2023a,Zhou2022a}. For overdoped cuprates, a pseudogap feature, possibly driven by phase fluctuations, has also been observed within a narrow region above $T_c$ \cite{Chen2022}. Such overall correspondence supports potential relevance of our work to the cuprate physics and highlights the importance of a strong-coupling real-space perspective in exploring high-temperature superconductivity.\\

\begin{figure*}[t]
\centering
\includegraphics[width=0.8\linewidth]{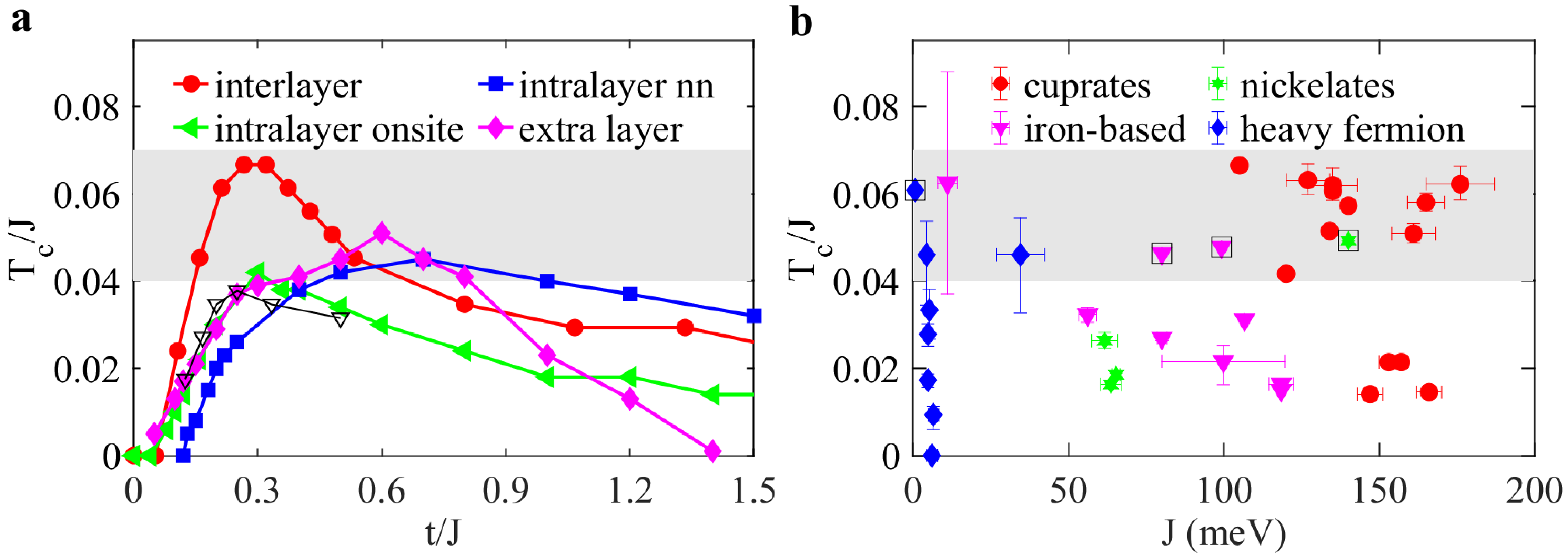}
\caption{\textbf{$\bm{T_c}/J$ ratio and its comparison with experiments.} \textbf{a} Typical results of $T_c/J$ as functions of $t/J$ for several effective models with interlayer, intralayer onsite, or intralayer nn (nearest-neighbour) pairing interactions. For simplicity, only the nearest-neighbour hopping $t$ is considered and the chemical potential is set to $\mu=0$. For nearest-neighbour pairing in a one-layer model (away from the van Hove singularity), introducing an extra layer of conduction electrons with nearest-neighbour interlayer hopping ($t_p=0.7t$) is found to enhance the maximum $T_c/J$. Also compared is a typical result of the attractive Hubbard model from previous quantum Monte Carlo simulations (open down-pointing triangles) away from the half-filling \cite{Paiva2010}. \textbf{b} Collection of experimental $T_c/J$ ratios for a number of cuprate, nickelate, iron-based, and heavy fermion superconductors, where $J$ are estimated from their respective spin interactions. The shaded area marks the region $T_c/J\approx 0.04-0.07$. The large error bar exceeding this region comes from bulk FeSe as discussed in the main text. All error bars come from the experimental uncertainty of $J$ as given by the original literatures listed in Table \ref{tab1}. The points circled by a square box mark those uncertain data from CePd$_2$Si$_2$, SmO$_{1-x}$F$_x$FeAs, LaO$_{1-x}$F$_x$FeAs, and La$_3$Ni$_2$O$_7$ discussed in the main text.}
	\label{fig5}
\end{figure*}

\begin{table*}
	\centering
\caption{Experimental data of the maximum $T_c$, the estimated paring interaction $J$, and the corresponding ratio $T_c/J$ in some of the cuprate, iron-based, nickelate, and heavy fermion superconductors. $J$ is the superexchange interaction derived mainly from RIXS for cuprate and nickelate superconductors and INS for iron-based superconductors. Most measurements on the latter only reported the value of $SJ$. Following the literature \cite{Zhao2009,Zhang2014,Ewings2011}, we have used the effective spin size $S=1/2$ to derive their $J$ except for $S=0.69$ in SrFe$_2$As$_2$. For bulk FeSe, the value of $T_c/J$ can be directly estimated from the literature with a large error bar \cite{}. Note that CePd$_2$Si$_2$, SmO$_{1-x}$F$_x$FeAs, LaO$_{1-x}$F$_x$FeAs, and La$_3$Ni$_2$O$_7$ are discussed in the main text but not included in the table due to the lack of unambiguous information on their $J$. For heavy fermion superconductors, $J$ is estimated crudely from the average coherence temperature. For simplicity, we refer to the original literatures for the errors of all listed data. \label{tab1}}
\begin{tabular}{lllllll}
	\hline
		&Nd$_{1-x}$Sr$_x$NiO$_2$	&Pr$_{1-x}$Sr$_x$NiO$_2$    &La$_{1-x}$Sr$_x$NiO$_2$  & CaFe$_2$As$_2$ 	&BaFe$_2$As$_2$&SrFe$_2$As$_2$   \\ \hline
		$T_c$(K)&12 \cite{Zeng2020a}  &14 \cite{Osada2020}   & 18.8 \cite{Sun2023d} & 25 \cite{Chen2016}&22.5 \cite{Kim2009}&21 \cite{Saha2009} \\ 
		$J$(meV)&63.6 \cite{Lu2021a}   & 66.5,\ 64 \cite{Gao2022} & 61.6 \cite{Rossi} & 99.8 \cite{Dai2015} &118.4\cite{Dai2015}
		&56.1 \cite{Dai2015}  \\ 
		$T_c/J$&0.016  &0.019   &0.026  &0.022 &0.016&0.032 \\
\hline
                           &\multicolumn{1}{l}{Ba$_{1-x}$K$_x$Fe$_2$As$_2$}   & BaFe$_{2-x}$Ni$_x$As$_2$     & \multicolumn{1}{l}{NaFeAs}            & \multicolumn{1}{l}{bulk FeSe}           & \multicolumn{1}{l}{CeCoIn$_5$}      & CeCu$_2$Si$_2$
                      \\ \hline
\multicolumn{1}{l}{$T_c$(K)}   & \multicolumn{1}{l}{38.5 \cite{Rotter2008a}}              & 20.5  \cite{Li2009}           & \multicolumn{1}{l}{25  \cite{Chu2009}}               & \multicolumn{1}{l}{8\cite{Hsu2008}}              & \multicolumn{1}{l}{2.3  \cite{Li2021b}}          & 0.7  \cite{Li2021b}          \\ 
\multicolumn{1}{l}{$J$(meV)}    & \multicolumn{1}{l}{106.6 \cite{Wang2013}}        & 118.4 \cite{Wang2013}         & \multicolumn{1}{l}{80 \cite{Dai2015}}                & \multicolumn{1}{l}{11.0 \cite{Gu2022}}          & \multicolumn{1}{l}{4.3 \cite{Yang2008b}}          & 6.5 \cite{Yang2008b}         \\ 
\multicolumn{1}{l}{$T_c/J$} & \multicolumn{1}{l}{0.031}          & 0.015            & \multicolumn{1}{l}{0.027}             & \multicolumn{1}{l}{0.063}         & \multicolumn{1}{l}{0.046}        & 0.0093       \\ \hline
\multicolumn{1}{l}{}     & \multicolumn{1}{l}{URu$_2$Si$_2$}         & UBe$_{13}$           & \multicolumn{1}{l}{UPd$_2$Al$_3$}           & \multicolumn{1}{l}{PuCoGa$_5$}        & \multicolumn{1}{l}{YbRh$_2$Si$_2$}     & YBa$_2$Cu$_4$O$_8$     \\ \hline
\multicolumn{1}{l}{$T_c$(K)}   & \multicolumn{1}{l}{1.5  \cite{Li2021b}}             & 0.95   \cite{Li2021b}           & \multicolumn{1}{l}{2  \cite{Li2021b}}                 & \multicolumn{1}{l}{18.4  \cite{Li2021b}}           & \multicolumn{1}{l}{0.002  \cite{Li2021b}}        & 81  \cite{Wang2022a}         \\ 
\multicolumn{1}{l}{$J$(meV)}    & \multicolumn{1}{l}{4.7 \cite{Yang2008b}}             & 4.7    \cite{Yang2008b}           & \multicolumn{1}{l}{5.2 \cite{Yang2008b}}               & \multicolumn{1}{l}{34.5 \cite{Pines2013}}           & \multicolumn{1}{l}{6.0 \cite{Yang2008b}}          & 105  \cite{Wang2022a}        \\ 
\multicolumn{1}{l}{$T_c/J$} & \multicolumn{1}{l}{0.028}           & 0.017             & \multicolumn{1}{l}{0.033}             & \multicolumn{1}{l}{0.046}          & \multicolumn{1}{l}{0.000028}     & 0.067        \\ \hline
\multicolumn{1}{l}{}     & \multicolumn{1}{l}{NdBa$_2$Cu$_3$O$_{6+\delta}$}    & Tl$_2$Ba$_2$CuO$_{6+\delta}$     & \multicolumn{1}{l}{HgBaCuO$_{4+\delta}$ }        & \multicolumn{1}{l}{HgBa$_2$CaCu$_2$O$_{6+\delta}$} & \multicolumn{1}{l}{La$_{2-x}$Sr$_x$CuO$_4$ } &Nd$_{2-x}$Ce$_x$CuO$_4$ \\ \hline
\multicolumn{1}{l}{$T_c$(K)}   & \multicolumn{1}{l}{95 \cite{Wang2022a}}              & 93    \cite{Wang2022a}            & \multicolumn{1}{l}{97 \cite{Wang2022a}}                & \multicolumn{1}{l}{127 \cite{Wang2022a}}            & \multicolumn{1}{l}{39 \cite{Wang2022a}}           & 24 \cite{Wang2022a}          \\
\multicolumn{1}{l}{$J$(meV)}    & \multicolumn{1}{l}{135 \cite{Wang2022a}}             & 127   \cite{Wang2022a}            & \multicolumn{1}{l}{135 \cite{Wang2022a}}               & \multicolumn{1}{l}{176 \cite{Wang2022a}}            & \multicolumn{1}{l}{157 \cite{Wang2022a}}          & 147   \cite{Wang2022a}       \\ 
\multicolumn{1}{l}{$T_c/J$} & \multicolumn{1}{l}{0.061}           & 0.063             & \multicolumn{1}{l}{0.062}             & \multicolumn{1}{l}{0.062}          & \multicolumn{1}{l}{0.021}        & 0.014        \\ \hline 
\multicolumn{1}{l}{}     & \multicolumn{1}{l}{Ca$_{2-x}$Na$_x$CuO$_2$Cl$_2$} & Bi$_2$Sr$_{2-x}$La$_x$CuO$_{6+\delta}$ & \multicolumn{2}{l}{Bi$_2$Sr$_{2-x}$La$_x$CuO$_{8+\delta}$} & \multicolumn{2}{l}{Bi$_{2+x}$Sr$_{2-x}$Ca$_2$Cu$_3$O$_{10+\delta}$}                                             \\ \hline
\multicolumn{1}{l}{$T_c$(K)}   & \multicolumn{1}{l}{28 \cite{Wang2022a}}              & 38    \cite{Wang2022a}            & \multicolumn{2}{l}{95 \cite{Wang2022a}}                & \multicolumn{2}{l}{111 \cite{Wang2022a}}                                                               \\ 
\multicolumn{1}{l}{$J$(meV)}    & \multicolumn{1}{l}{166 \cite{Wang2022a}}             & 153  \cite{Wang2022a}             & \multicolumn{2}{l}{161 \cite{Wang2022a}}               & \multicolumn{2}{l}{165 \cite{Wang2022a}}                                                               \\ 
\multicolumn{1}{l}{$T_c/J$} & \multicolumn{1}{l}{0.015}           & 0.021             & \multicolumn{2}{l}{0.051}             & \multicolumn{2}{l}{0.058}                                                             \\ \hline
\multicolumn{1}{l}{}     & \multicolumn{2}{l}{(Ca$_{0.1}$La$_{0.9}$)(Ba$_{1.65}$La$_{0.35}$)Cu$_3$O$_y$}     & \multicolumn{2}{l}{(Ca$_{0.4}$La$_{0.6}$)(Ba$_{1.35}$La$_{0.65}$)Cu$_3$O$_y$}                         & \multicolumn{1}{l}{}             &              \\ \hline
\multicolumn{1}{l}{$T_c$(K)}   & \multicolumn{2}{l}{58 \cite{Ofer2006a}}                                  & \multicolumn{2}{l}{80 \cite{Ofer2006a}}                                                      & \multicolumn{1}{l}{}             &              \\
\multicolumn{1}{l}{$J$(meV)}    & \multicolumn{2}{l}{120 \cite{Ellis2015}}                                 & \multicolumn{2}{l}{134 \cite{Ellis2015}}                                                     & \multicolumn{1}{l}{}             &              \\ 
\multicolumn{1}{l}{$T_c/J$} & \multicolumn{2}{l}{0.042}                               & \multicolumn{2}{l}{0.052}                                                   & \multicolumn{1}{l}{}             &              \\ \hline
\end{tabular}
\end{table*}

\noindent\textbf{Constraint on maximum $T_c/J$}\\
Another important observation of our calculations is that the superconductivity may intrinsically be suppressed for sufficiently strong pairing interaction even without considering competing orders from other channels. Thus, $T_c$ is constrained from both sides of strong and weak pairing interactions. It is then sensible to study the ratio $T_c/J$ to have a feeling about the maximum $T_c$ allowed by the pairing interaction $J$ \cite{Wang2022a,Wang2023a,Ruan2016}. For the one band square lattice model discussed so far, we find a maximum ratio $T_c/J\approx 0.04$. We have also tested other parameters and find that tuning the next-nearest-neighbour hopping $t'$ and the chemical potential $\mu$ can only slight improve this ratio. Specifically, at half-filling with $t'=0$ and $\mu=0$ near the van Hove singularity, the maximum $T_c/J$ may be enhanced to 0.045. Motivated by the possible importance of apical oxygen on $T_c$ \cite{Scalapino2012a}, we have also studied a model with an extra conduction layer, and find the maximum $T_c/J$ may be at most enhanced to about $0.06$ for certain special (nearest-neighbour) interlayer hopping. On the other hand, local interlayer hopping is found to suppress this maximum ratio. Taking $t\approx 100-200$ meV from the angle-resolved photoemission spectroscopy (ARPES) and the specific heat analysis \cite{Chen2022,Harrison2023} and $J\approx 100-190$ meV from the resonant inelastic X-ray scattering measurement (RIXS) \cite{Wang2022a,Ellis2015}, these ratios yield the highest $T_c$ to be 100-130 K, consistent with the reported $T^{\rm max}_c=$97 K for single-layer and $135$ K for multi-layer cuprate superconductors under ambient pressure \cite{Scalapino2012a,Antipov2002}.

To further explore the above idea, we extend our calculations to other variations of the minimum effective model, covering nearest-neighbour or onsite pairings, single or multi-layer structures, and intralayer or interlayer pairings (see Methods). It is important to note that our models do not depend on fine details of the microscopic pairing mechanism, as long as the effective pairing interaction and the low-energy Hamiltonians remain the same. Figure \ref{fig5}\textbf{a} shows the variations of $T_c/J$ versus $t/J$ in these models, where $J$ is the local attractive Hubbard interaction for onsite pairing, and the interlayer superexchange interaction for interlayer pairing as discussed previously for La$_3$Ni$_2$O$_7$ under high pressure \cite{Yang2023, Qin2023a}. We see all curves behave nonmonotonically with the pairing interaction, although they may have different strong-coupling limit (e.g., BEC for onsite pairing and preformed local interlayer pairing for bilayer nickelates), with the maximum $T_c/J$ lying within the interval from 0.04 to 0.07. Notably, for the attractive Hubbard model, our simulations yield consistent results compared with previous quantum Monte Carlo simulations (open down-pointing triangles) \cite{Paiva2010}, which reinforces the reliability of our approach, and introducing an additional conduction layer gives a similar maximum ratio \cite{Wachtel2012,Berg2008PRB}. Note that we have ignored long distance pairing since it is typically weaker than onsite or nearest-neighbour ones for reaching the maximum $T_c$. All other instabilities are also ignored to maximize the pairing instability. Hence our phase diagram is not the full phase diagram with all possible ground states of a physical model, but a phase diagram that intentionally exaggerates the superconductivity and other possible instabilities in the pairing channel, so that the derived $T_c/J$ could be a better estimate of its potential upper limit.

To see if the above constraint may indeed apply in real materials, Fig.~\ref{fig5}\textbf{b} and Table \ref{tab1} collect the data for a number of well-known unconventional superconductors \cite{Zhao2009,Zhang2014,Ewings2011,Zeng2020a,Osada2020,Sun2023d,Chen2016,Kim2009,Saha2009,Lu2021a,Gao2022,Rossi,Dai2015,Rotter2008a,Li2009,Chu2009,Hsu2008,Li2021b,Wang2013,Gu2022,Yang2008b,Wang2022a,Pines2013,Ofer2006a,Ellis2015}. The spin energy scale in cuprate, iron-based, and nickelate superconductors have been determined mainly by the spin wave fitting in inelastic neutron scattering (INS) or RIXS experiments \cite{Ofer2006a,Tacon2011,Dean2013,Wang2013,Pelliciari2016}, where $J$ has been found to vary only slightly with doping. It differs from the renormalized one due to the feedback effect observed in low-energy measurements by INS \cite{Wakimoto2004} and two-magnon extraction in Raman spectra \cite{Li2012}. In iron-based superconductors such as CaFe$_2$As$_2$, SrFe$_2$As$_2$, BaFe$_2$As$_2$, and NaFeAs, the ratios $T_c/J$ are less than 0.063 \cite{Dai2015,Chen2016,Chu2009,Wang2013,Gu2022,Kim2009,Saha2009,Rotter2008a,Li2009,Hsu2008}, where the value of $J$ is extracted from the reported $SJ$ by taking the effective spin size $S=0.69$ for SrFe$_2$As$_2$ and $S=1/2$ for all others except for FeSe following the literatures \cite{Zhao2009,Zhang2014,Ewings2011}. The large error bar exceeding the shaded area in Fig.~\ref{fig5}\textbf{b} comes from the bulk FeSe ($T_c=8$ K), for which neutron scattering measurements reported the ratio $T/J=0.86\pm 0.35$ at $T=110$ K \cite{Gu2022}. Unfortunately, we do not find the data for FeSe films, whose high $T_c$ might involve contributions from the interface. To the best of our knowledge, there is also no exact estimate of $J$ for the 1111 systems. It has been reported that SmOFeAs adopts an intermediate spin dispersion between those of NaFeAs and BaFe$_2$As$_2$ \cite{Pelliciari2016a}. Assuming that the spin interaction is not sensitive to the doping, as observed in BaFe$_{2-x}$Ni$_x$As$_2$ and NaFe$_{1-x}$Co$_x$As \cite{Wang2013,Pelliciari2016}, we might roughly estimate $J\sim80-118.4$ meV for SmO$_{1-x}$F$_x$FeAs and thus obtain a maximum ratio $T_c/J \approx 0.040-0.059$ given its maximum $T_c$=55 K \cite{Ren2008}. While for LaO$_{1-x}$F$_x$FeAs, experiments only indicate an overall magnitude of $SJ\sim40$ meV along different directions \cite{Ramazanoglu2013}, which yields $T_c/J\sim0.046$ with its maximum $T_c=43$ K using $S=1/2$ \cite{DelaCruz2008}. Both fall within our proposed range.

The infinite-layer nickelate superconductors have a small maximum ratio of about 0.026, which indicates the potential to reach a higher $T_c$ \cite{Zeng2020a,Osada2020,Lu2021a,Gao2022,Sun2023d,Rossi,Chow2024}. RIXS measurements \cite{Chen2024} on the high-pressure high-temperature bilayer nickelate superconductor La$_3$Ni$_2$O$_7$ reported an interlayer spin interaction strength ($J$) of about 140 meV assuming its spin size $S=1/2$, which also seems to be confirmed by inelastic neutron measurements \cite{Xie2024}. Although these measurements were performed under ambient pressure, it gives a rough estimate of the magnitude of $J$. If we naively apply this value to the high pressure region where the superconductivity was reported with $T_c^{\rm max}\approx80$ K, we find $T_c^{\rm max}/J\approx 0.05$ for the bilayer nickelate superconductors, which agrees well with our previous Monte Carlo simulations \cite{Qin2023a}. Recently, superconductivity has been reported also in the trilayer nickelate superconductor La$_4$Ni$_3$O$_{10}$ under high pressure, albeit with a much smaller $T_c^{\rm max}\approx 30$ K \cite{Zhu2024}. It has been proposed theoretically that competition and frustration of interlayer pairing between the inner layer and two outer layers may lead to strong superconducting fluctuations and thus reduce the maximum ratio of $T_c/J$ to $0.02-0.03$ \cite{Qin2024}. This, together with layer imbalance and the possibly smaller interlayer $J$, may explain the much reduced $T_c^{\rm max}$ in the trilayer nickelate compared to those in the bilayer ones. 

By contrast, the cuprate high-temperature superconductors have the highest $T_c^{\rm max}$ in the trilayer structure, and their overall $T_c^{\rm max}/J$ ratios can reach up to 0.067, as observed in HgBa$_2$CaCu$_2$O$_{6+\delta}$, YBa$_2$Cu$_4$O$_8$, YBa$_2$Cu$_3$O$_{6+\delta}$, NdBa$_2$Cu$_3$O$_{6+\delta}$, Tl$_2$Ba$_2$CuO$_{6+\delta}$, and Bi$_{2+x}$Sr$_{2-x}$Ca$_2$Cu$_3$O$_{10+\delta}$ \cite{Wang2022a}. This opposite tendency reflects an intrinsic distinction in the pairing mechanisms between multilayer nickelate and cuprate superconductors \cite{Yang2023,Luo2024,Qin2024,Qin2023a}. In heavy-fermion superconductors such as CeCoIn$_5$ or PuCoGa$_5$, systematic measurements of $J$ are lacking. We therefore estimate the spin interaction energy from the coherence temperature scale, namely the Ruderman-Kittle-Kasuya-Yosida (RKKY) scale, and find the highest $T_c/J$ to be about $0.046$ \cite{Yang2008b,Li2021b,Pines2013}. To the best of our knowledge, a spin wave fitting has only been applied to CePd$_2$Si$_2$ and yields $J=0.61$ meV under ambient pressure \cite{Dijk2000}. Combining naively this value with its $T_c=0.43$ K at 3 GPa gives the ratio $T_c/J\approx 0.061$, in good alignment with our suggested constraint.

Despite vast complexities across all these different families of unconventional superconductors that are far beyond our simplified models, none of their maximum $T_c/J$ ratios exceeds the proposed range $0.04-0.07$, suggesting that our calculations indeed capture some essence of the fundamental physics of unconventional superconductivity. While this range of the maximum $T_c/J$ ratio is also supported by data collection in previous studies \cite{Ofer2006a,Moreira2001,Grzelak2020}, one should keep in mind that our calculations only cover a small part of the parameter and model space, and it is not completely clear how other complex factors, such as the dimensionality and multi-orbitals, might affect the ratio. The maximum ratio proposed here seems to more represent a practical upper limit for some quite generic situations in real superconductors. This consistency urges for a more rigorous theoretical understanding. 

We finally comment on the $T_c/t$ ratio widely used in previous literatures. Unlike $T_c/J$, we find the maximum $T_c/t$ depends more sensitively on models and may reach 0.29, 0.15, 0.105 upon tuning the hopping parameters or the chemical potential for interlayer, intralayer onsite, intralayer nearest-neighbour (nn) pairings, respectively. For the attractive Hubbard model, our derived maximum $T_c/t\approx 0.15$ is close to the quantum Monte Carlo result of $0.17$, which also confirms the validity of our estimate \cite{Paiva2010}. In general, our calculations across these different models yield the maximum $T_c/t$ ratio within the range of $0.1-0.3$ at an optimal $t/J$ value of around 0.2, somewhat different from those for the maximum $T_c/J$ ratio. The relatively lager variation of the $T_c/t$ ratio may be ascribed to the fact that the long-range phase coherence determining $T_c$ relies heavily on the cooperative hopping of paired electrons and may hence differ greatly for different pairing configurations and lattice geometries beyond the simple hopping parameters. The quasiparticle hopping is also strongly renormalized by correlation effects, which makes it difficult to estimate in practice. It is for these reasons that we have chosen to treat $t$ as a tuning parameter and focus on the $T_c/J$ ratio that might be better compared with experiment.\\

\noindent\textbf{Route to room temperature superconductivity?}\\
It is important to emphasize again that the above agreement by no means implies that all these superconductors, including hole-doped cuprates, are fully described by the specified pairing mechanisms in our simplified models. There is also no rigorous theoretical proof for a maximum $T_c$ in unconventional superconductors \cite{Hofmann2022}. Nevertheless, if we take the above constraint seriously, achieving room temperature superconductivity seems unlikely under ambient pressure within the current theoretical framework. For $T_c$ to reach 300 K, we need a pairing interaction of the order $400-700$ meV, which is twice higher than the spin exchange interaction in cuprates and seems unrealistic based on our survey of existing correlated materials. Moreover, the maximum $T_c/J$ is only realized at an optimal ratio of $t/J$, thus also requiring a larger quasiparticle hopping $t$, a situation that seems to only occur under pressure. This is contrary to the weak-coupling BCS theory which predicts a higher $T_c$ for a larger density of states (smaller $t$), while in the strong coupling limit, the pairing strength is sufficiently large and one requires sufficient kinetic energy, hence a larger $t$, to achieve the phase coherence. For instance, in the attractive Hubbard model with an infinite $U$, $T_c$ is determined by a small fraction of $t$. Consequently, for room-temperature superconductivity, $t$ must reach the order of hundred meV, which is not favored in flat-band systems \cite{Cao2018}.

 It is therefore imperative to explore alternative avenues to enhance the ratio under ambient pressure. It has been noticed that three-layer cuprate superconductors have the highest $T_c$. One may therefore speculate that multi-layer may promote $T_c$. Indeed, the maximum $T_c$ increases from 97 K in the single-layer HgBa$_2$CuO$_{4+\delta}$ to 127 K in the two-layer HgBa$_2$CaCu$_2$O$_{6+\delta}$ and 135 K in the three-layer HgBa$_2$Ca$_2$Cu$_3$O$_{9+\delta}$ \cite{Scalapino2012a,Antipov2002}. However, the ratio $T_c/J\approx 0.062$ seems to remain unchanged and the increase of $T_c$ seems to come purely from the increase of $J$ \cite{Wang2022a}. On the other hand, the maximum $T_c/J$ does increase from 0.021 in the single-layer Bi$_2$Sr$_{2-x}$La$_x$CuO$_{6+\delta}$ to 0.058 in the three-layer Bi$_{2+x}$Sr$_{2-x}$Ca$_2$Cu$_3$O$_{10+\delta}$ in Bi-systems \cite{Wang2022a}, but the latter still lies within our proposed range, implying that increasing the number of layers from Bi2201 to Bi2223 only helps to tune the optimal conditions for maximizing $T_c/J$, while the constraint itself is not touched. We have also examined the effect of additional local interlayer hopping and find that it actually reduces the maximum $T_c/J$. Additionally, one may follow the studies of FeSe films \cite{Song2019,Lee2014a} and consider to improve $T_c$ by introducing phonons, but this seems empirically at most to provide an increase of around 40 K, given the limited characteristic phonon frequencies under ambient pressure \cite{Cai2023,McMillan1968}. A larger spin interaction occurs for the Hund's rule coupling inside an atom. However, it is not clear if intra-atomic inter-orbital pairing may support a high $T_c$ due to their very different orbital characters of the paired electrons. 

\vspace{20pt}
\noindent
\textbf{Discussion}\\
Taken together, the known unconventional superconductor families seem to have almost exhausted their potentials in reaching the highest $T_c$ allowed by their spin exchange interactions. As a result, room-temperature superconductivity at ambient pressure seems unlikely to arise from a single pairing mechanism within the current theoretical framework, unless one could find a way to substantially enhance the exchange interaction. Our results may not only help rule out some evidently wrong directions \cite{Lee2023a,Lee2023b}, but also point out the necessity of exploring alternative approaches to achieve room-temperature superconductors at ambient pressure \cite{NP2006,Berg2008PRB,Xiang2015P,Hu2016SciB,Pines2016RPP,Yang2023,Pines2013,Zhang2023,Lee2018,Kivelson2020,Dahm2009,Basov2011,Volovik2018}. It encourages the possibility of incorporating different pairing mechanisms \cite{Lederer2015,Dzhumanov1996,Barantani2022,Monthoux2007,Sous2018,Aji2010}, including but not limited to magnetic, charge, orbital fluctuations, or excitons, bipolarons, etc, to improve the overall effective pairing interaction, for which FeSe films may be a good example \cite{Wang2012CPL,Liu2012,Tan2013}. Our derived ratio provides a tentative guide for future material exploration of novel high-temperature superconductors. Theoretically, by utilizing $J$ from newly developed methods \cite{Cui2022} and effective hopping $t$ from strongly correlated calculations \cite{DiCataldo2024}, an approximate estimate of the upper limit of $T_c$ may be predicted for the selection of promising candidates. Experimentally, estimating $J$ from RIXS, INS, or other state-of-the-art techniques in newly discovered materials may also help identify their potential in reaching the desired $T_c$. Last but not least, understanding unconventional superconductivity from a real-space, strong-coupling perspective may already provide an operational and more practical avenue for material design compared to the momentum-space, weak-coupling approach.

\vspace{20pt}
\noindent
\textbf{Methods}

{\noindent \textbf{Pairing interaction}}\\
For onsite pairing in the attractive Hubbard model, we have immediately
	\begin{eqnarray}
	-U\sum_{i}d_{i\uparrow}^{\dagger}d_{i\uparrow}d_{i\downarrow}^{\dagger}d_{i\downarrow}=J\sum_{i}\psi_{i}^{\dagger}\psi_{i},
	\end{eqnarray}
	where $\psi_{i}=d_{i\downarrow}d_{i\uparrow}$ and $J=U$.

For the Hubbard model with a large Coulomb repulsion $U$, the pairing interaction $J$ is given by the exchange interaction between nearest-neighbor sites. To see this, we may follow the standard derivation to first project out the double occupancy and map the Hubbard model to an effective low-energy model with the following interaction term:
	\begin{eqnarray}
		V_{\rm int}=J_{\rm ex}\sum_{\langle ij\rangle}\left(\bm{s}_i\cdot \bm{s}_j-\frac{1}{4}n_in_j\right),
	\end{eqnarray} 
	where $\bm{s}_{i}=\sum_{ \alpha,\beta}d_{i\alpha}^{\dagger}\frac{\bm{\sigma}_{\alpha\beta}}{2}d_{i\beta}$ is the spin density and $n_i=\sum_{\sigma}d_{i\sigma}^{\dagger}d_{i\sigma}$ is the charge density.

To study the superconductivity, we introduce the spin-singlet ($\psi_{ij}^{\rm S}$) and spin-triplet ($\bm{\psi}_{ij}^{\rm T}$) operators: 
	\begin{equation}
		\begin{split}
			\psi_{ij}^{\rm S}&=\frac{1}{\sqrt{2}}\sum_{ \alpha,\beta}d_{i\alpha}(-i\sigma_y)_{\alpha\beta}d_{j\beta},\\
			\bm{\psi}^{\rm T}_{ij}&=\frac{1}{\sqrt{2}}\sum_{ \alpha,\beta}d_{i\alpha}(-i\sigma_y \bm{\sigma})_{\alpha\beta}d_{j\beta},
		\end{split}
	\end{equation}
where $\psi_{ij}^{\rm S}$ and $\bm{\psi}_{ij}^{\rm T}$ satisfy $\psi_{ji}^{\rm S}=\psi_{ij}^{\rm S}$ and $\bm{\psi}_{ji}^{\rm T}=-\bm{\psi}_{ij}^{\rm T}$.

The above interaction can then be rewritten as:
	\begin{eqnarray}
		V_{\rm int}=-J_{\rm ex}\sum_{\langle ij\rangle}\left(\psi_{ij}^{\rm S}\right)^{\dagger}\psi_{ij}^{\rm S},
	\end{eqnarray}
where the spin-triplet term cancels. This is the Hamiltonian (with $J=J_{\rm ex}$) used in our simulations for the $d$-wave superconductivity. For cuprates, $J$ is given by the superexchange mechanism. Note that to maximize $T_c$, we have included both the nearest-neighbor antiferromagnetic spin-density interaction, $J \bm{s}_i \cdot \bm{s}_j$, and the nearest-neighbor attractive charge-density interaction, $-V^c n_i n_j$. While only the former is included in many works, both seem to be supported by recent experiments \cite{Chen2021Science,OMahony2022}. For spin-fluctuation interaction \cite{Wang2022a,Monthoux2007}, $V(\bm{q}) \bm{s}_{\bm{q}} \cdot \bm{s}_{-\bm{q}}$, the dominant contribution has a similar form in real space. These justify our choice of the phenomenological pairing term.
	
Similarly, we have for interlayer pairing: 
	\begin{eqnarray}
	J_{\rm ex}\sum_{i}\left(\bm{s}_{1i}\cdot\bm{s}_{2i}-\frac{1}{4}n_{1i}n_{2i}\right)=-J\sum_{i}\psi_{12i}^{\dagger}\psi_{12i}.
	\end{eqnarray}
	where $\bm{s}_{ai}=\sum_{ \alpha,\beta}d_{ai\alpha}^{\dagger}\frac{\bm{\sigma}_{\alpha,\beta}}{2}d_{aj\beta}$, $n_i=\sum_{\sigma}d_{ai\sigma}^{\dagger}d_{ai\sigma}$, and $\psi_{12i}=\frac{1}{\sqrt{2}}(d_{1i\downarrow}d_{2i\uparrow}-d_{1i\uparrow}d_{2i\downarrow})$. Again, the pairing interaction is $J=J_{\rm ex}$.
	\\

{\color{black}\noindent \textbf{Mutual information and vortex number}}\\
The superconducting transition temperature is always determined by the superconducting phase coherence from the phase mutual information defined as \cite{Qin2023PRB,Qin2023a}:
\begin{equation}
I^{x/y}_{\bm R}=\int d\theta^{x/y}_{\bm{0}} d\theta^{x/y}_{\bm{R}}~ p(\theta^{x/y}_{\bm{0}},\theta^{x/y}_{\bm{R}})\ln\frac{p(\theta^{x/y}_{\bm{0}},\theta^{x/y}_{\bm{R}})}{p(\theta^{x/y}_{\bm{0}})p(\theta^{x/y}_{\bm{R}})},
\end{equation}
where $p(\theta_{\bm{0}}^{x/y}),~p(\theta_{\bm{R}}^{x/y})$ is the marginal distribution of the pairing field phase on two bonds with a distance $\bm{R}$, and $p(\theta_{\bm{0}},\theta_{\bm{R}})$ is their joint probabilistic distribution. For onsite or interlayer pairing,  $\theta_{\bm{R}}^{x/y}$ simplifies to $\theta_{\bm{R}}$. 

The vortex number is calculated using
\begin{equation}
n_{\rm v}=\sum_{i}\langle \delta_{w_i,1}\rangle,
\end{equation}
where $w_i$ is the winding number for $\theta_i\rightarrow\theta_{i+\hat{x}}\rightarrow\theta_{i+\hat{x}+\hat{y}}\rightarrow\theta_{i+\hat{y}}\rightarrow\theta_i$ with the phase $\theta_i$ of $\Delta_i=(\Delta_{i,i+\hat{x}}+\Delta_{i,i-\hat{x}}-\Delta_{i,i+\hat{y}}-\Delta_{i,i-\hat{y}})/4$  for nearest-neighbour pairing and $\langle \rangle$ denotes the statistic average over all pairing configurations. For onsite or interlayer pairing, $\theta_i$ is the phase of the pairing field at site $i$. 

We find that $T_c$ determined from the phase mutual information agrees well with that estimated from the numerical derivatives of the vortex number $\frac{dn_{\rm v}}{dT}$ as well as the BKT transition temperature obtained from the superfluid stiffness \cite{Nelson1977}, which confirms the validity of our method in calculating $T_c$ for these models. However, the superfluid stiffness is computationally more expensive. \\

\begin{figure}[t]
	\centering
	\includegraphics[width=8cm]{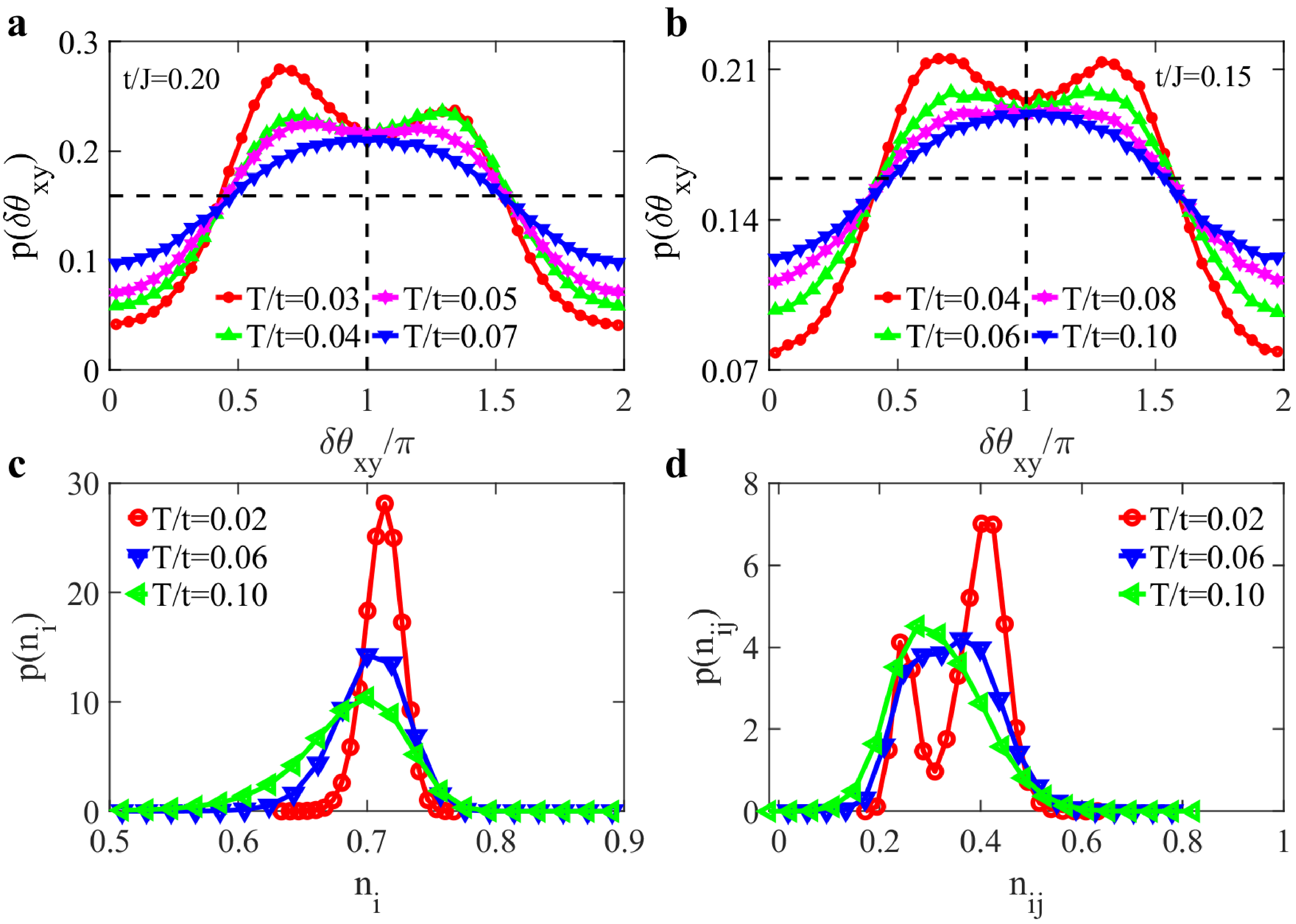}
	\caption{\textbf{Coexisting time reversal symmetry breaking and bond charge order of the plaquette state.} \textbf{a,b} Probabilistic distributions of the phase difference $\delta \theta_{xy}$ along $x$ and $y$ direction for $t/J=0.20$ and 0.15, respectively. \textbf{c,d} Comparison of the distribution of the site electron density $n_i$ and the bond charge density $n_{ij}$ for $t/J=0.15$.
}
	\label{fig6}
\end{figure}

{\noindent \textbf{Time-reversal symmetry breaking and coexisting bond charge order of the plaquette state}}\\
Figures \ref{fig6}\textbf{a} and \ref{fig6}\textbf{b} plot the probabilistic distributions of the phase difference $\delta \theta_{xy}$ along $x$ and $y$ directions for $t/J = 0.20$ and 0.15, where the plaquette state develops at low temperatures. We see two-peak structures below roughly $T=0.06$ for $t/J = 0.20$ and $T=0.09$ for $t/J = 0.15$, both of which coincide with $T_{\scriptscriptstyle \square}$ determined in Fig. \ref{fig1}\textbf{a} from the distribution of the pairing amplitude and other quantities. As discussed in the main text, the deviation from $\delta\theta_{xy}=\pi$ marks the time-reversal symmetry breaking. Figures \ref{fig6}\textbf{c} and \ref{fig6}\textbf{d} further compare the distributions of the site charge density $n_i = \sum_{\sigma} \langle d_{i\sigma}^{\dagger} d_{i\sigma} \rangle$ and the bond charge density $n_{ij} = \sum_{\sigma} \langle d_{i\sigma}^{\dagger} d_{j\sigma} + d_{j\sigma}^{\dagger} d_{i\sigma} \rangle$  between two nearest-neighbour sites. While the former always exhibits a single peak, the latter also develops a two-peak structure below $T_{\scriptscriptstyle \square}$. These indicate a uniform charge distribution on all sites but spatial modulation of the bond charge density. We thus conclude that the plaquette state below $T_{\scriptscriptstyle \square}$ breaks simultaneously the time-reversal symmetry and the translational symmetry and coexists with a bond charge order.\\

{\color{black}\noindent \textbf{Other models}}\\
To derive the $T_c/J$ constraint, we extend the simplest one-band model to the following variations:

(1) A two-layer model with intralayer nearest-neighbour pairing and interlayer hopping:
\begin{eqnarray}
H&=&-\sum_{aij,\sigma}t_{ij}d_{ai\sigma}^{\dagger}d_{aj\sigma}-\mu\sum_{ai\sigma}d_{ai\sigma}^{\dagger}d_{ai\sigma}\\
&&-J\sum_{a\langle ij\rangle}\psi_{aij}^{\dagger}\psi_{aij}
-\sum_{i\sigma}t_p(d_{1i\sigma}^{\dagger}d_{2i\sigma}+h.c.),\nonumber
\end{eqnarray}
where the subscript $a=1,2$ represents the layer index, $\psi_{aij}=\frac{1}{\sqrt{2}}(d_{ai\downarrow}d_{aj\uparrow}-d_{ai\uparrow}d_{aj\downarrow})$, and $t_p$ denotes the local interlayer hopping.

(2) A two-layer model with an extra conduction layer motivated by the possible importance of apex oxygens in cuprates:
\begin{eqnarray}
H&=&-\sum_{ij,\sigma}t_{ij}d_{i\sigma}^{\dagger}d_{j\sigma}-\mu\sum_{i\sigma}d_{i\sigma}^{\dagger}d_{i\sigma}-J\sum_{\langle ij\rangle}\psi_{ij}^{\dagger}\psi_{ij}\nonumber\\
&&-\sum_{ij,\sigma}t_{ij}^cc_{i\sigma}^{\dagger}c_{j\sigma}-\mu^c\sum_{i\sigma}c_{i\sigma}^{\dagger}c_{i\sigma}\nonumber\\
&&-\sum_{ij,\sigma}t_p(c_{i\sigma}^{\dagger}d_{j\sigma}+h.c.),
\end{eqnarray}
where $t_p$ denotes local ($i=j$) or nearest-neighbour ($j=i\pm\hat{x}$ or $i\pm\hat{y}$) interlayer hopping.

(3) A single-layer model with onsite pairing interaction as in the attractive Hubbard model:
\begin{eqnarray}
H=-\sum_{ij,\sigma }t_{ij}d_{i\sigma}^{\dagger}d_{j\sigma}-J\sum_{i}\psi_{i}^{\dagger}\psi_{i},
\end{eqnarray}
where $\psi_{i}=d_{i\downarrow}d_{i\uparrow}$ and $J$ is given by the local attractive Hubbard interaction.

(4) A two-layer model with onsite pairing in one layer and an extra conduction layer:
\begin{eqnarray}
H&=&-J\sum_{i}\psi_{i}^{\dagger}\psi_{i}-\sum_{ij,\sigma}t_{p}(c_{i\sigma}^{\dagger}d_{j\sigma}+h.c.)\nonumber\\
&&-\sum_{ij,\sigma }t^c_{ij}c_{i\sigma}^{\dagger}c_{j\sigma}-\mu^c\sum_{i\sigma}c_{i\sigma}^{\dagger}c_{i\sigma},
\end{eqnarray}
where $\psi_{i}=d_{i\downarrow}d_{i\uparrow}$ and $t_p$ denotes local ($i=j$) or nearest-neighbour ($j=i\pm\hat{x}$ or $i\pm\hat{y}$) interlayer hopping.

(5) A two-layer model with interlayer pairing:
\begin{eqnarray}
H=-\sum_{aij,\sigma }t_{ij}d_{ai\sigma}^{\dagger}d_{aj\sigma}-J\sum_{i}\psi_{12i}^{\dagger}\psi_{12i},
\end{eqnarray}
where $\psi_{12i}=\frac{1}{\sqrt{2}}(d_{1i\downarrow}d_{2i\uparrow}-d_{1i\uparrow}d_{2i\downarrow})$.

Models (1) and (2) are constructed to reflect the effects of interlayer hopping and apical oxygen in cuprate superconductors. Model (2) may also be applied to the infinite-layer nickelate superconductors. Models (3) and (4) deal with onsite pairing with local attractive interaction. And model (5) is motivated by the bilayer nickelate superconductor.\\

\vspace{20pt}
\noindent    
\textbf{Data availability}
\noindent The data that support the findings of this study are available from the corresponding author upon reasonable request.

\vspace{20pt}
\noindent  
\textbf{Code availability}
\noindent The code in this study are available from the corresponding author upon reasonable request.

\vspace{20pt}
\noindent        
\textbf{Acknowledgements}

\noindent This work was supported by the National Natural Science Foundation of China (Grants No. 12474136 and No. 12174429), the Strategic Priority Research Program of the Chinese Academy of Sciences (Grant No. XDB33010100), and the National Key R\&D Program of China (Grant No. 2022YFA1402203).

\vspace{20pt}
\noindent
\textbf{Author contributions}

\noindent Y.Y. conceived the idea and supervised the project;  Q.Q. performed the calculations; Y.Y. and Q.Q. wrote the paper.

\vspace{20pt}
\noindent
\textbf{Competing interests}

\noindent The authors declare no competing interests.

\vspace{20pt}
\noindent
\textbf{Additional information}

\noindent \textbf{Correspondence} and requests for materials should be addressed to Y.Y.

\end{document}